\documentclass[letterpaper, 10pt, conference]{ieeeconf}
\IEEEoverridecommandlockouts
\usepackage{amsmath,amsfonts,bm}
\usepackage{graphicx}

\title{\LARGE \bf
Improving Convergence of Belief Propagation Decoding}

\author{M.G. Stepanov and M. Chertkov %
 \thanks{This work was carried out under the auspices of the National
 Nuclear Security Administration of the
 U.S. Department of Energy at Los Alamos National Laboratory under
 Contract No. DE-AC52-06NA25396.}
\thanks{M.G.~Stepanov is with Theoretical Division and Center for
Nonlinear Studies, LANL, Los Alamos, NM 87545, USA and Institute of
Automation and Electrometry, Novosibirsk 630090, Russia (on leave);
{\tt\small stepanov@cnls.lanl.gov}}%
\thanks{M.~Chertkov is with Theoretical Division and Center for
Nonlinear Studies, LANL, Los Alamos, NM 87545, USA;
{\tt\small chertkov@lanl.gov}}%
}

\begin{document}

\maketitle
\thispagestyle{empty}
\pagestyle{empty}

\begin{abstract}
The decoding of Low-Density Parity-Check codes by the Belief Propagation
(BP) algorithm is revisited. We check the iterative algorithm for its
convergence to a codeword (termination), we run Monte Carlo
simulations to find the probability distribution function of the
termination time, $n_{\rm it}$. Tested on an example $[155, 64,
20]$ code, this termination curve shows a maximum and an extended
algebraic tail at the highest values of $n_{\rm it}$. Aiming to
reduce the tail of the termination curve we consider a family of
iterative algorithms modifying the standard BP by means of a simple
relaxation. The relaxation parameter controls the convergence of the
modified BP algorithm to a minimum of the Bethe free energy. The
improvement is experimentally demonstrated for
Additive-White-Gaussian-Noise channel in some range of the
signal-to-noise ratios. We also discuss the trade-off between
the relaxation parameter of the improved iterative scheme and the number
of iterations.
\end{abstract}

Low-Density Parity-Check (LDPC) codes \cite{63Gal,99Mac} are the best
linear block error-correction codes known today \cite{03RU}. In addition
to being good codes, i.e. capable of decoding without errors in the
thermodynamic limit of an infinitely long block length, these codes can
also be decoded efficiently. The main idea of Belief Propagation (BP)
decoding is in approximating the actual graphical model, formulated for
solving statistical inference Maximum Likelihood (ML) or
Maximum-A-Posteriori (MAP) problems, by a tree-like structure without
loops. Being efficient but suboptimal the BP algorithm fails on certain
configurations of the channel noise when close to optimal (but
inefficient) MAP decoding would be successful.

BP decoding allows a certain duality in interpretation. First of all,
and following the so-called Bethe-free energy variational approach
\cite{05YFW}, BP can be understood as a set of equations for beliefs
(BP-equations) solving a constrained minimization problem. On the
other hand, a more traditional approach is to interpret BP in terms
of an iterative procedure --- so-called BP iterative algorithm
\cite{63Gal,88Pea,99Mac}. Being identical on a tree (as then BP
equations are solved explicitly by iterations from leaves to the
tree center) the two approaches are however distinct for a graphical
problem with loops. In case of their convergence, BP algorithms find
a minimum of the Bethe free energy \cite{05YFW,02TJ,04Hes},  however
in a general case convergence of the standard iterative BP is not
guaranteed. It is also understood that BP fails to converge
primarily due to circling of messages in the process of iterations
over the loopy graph.

To enforce convergence of the iterative algorithm to a minimum of
the Bethe Free energy some number of modifications of the standard
iterative BP were discussed in recent years. The tree-based
re-parametrization framework of \cite{03WJW} suggests to limit
communication on the loopy graph, cutting some edges in a dynamical
fashion so that the undesirable effects of circles are suppressed.
Another, so-called concave-convex procedure, introduced in
\cite{02Yui} and generalized in \cite{03HAK}, suggests to decompose
the Bethe free energy into concave and convex parts thus splitting
the iterations into two sequential sub-steps.

Noticing that convergence of the standard BP fails mainly due to
overshooting of iterations,  we develop in this paper a tunable
relaxation (damping) that cures the problem. Compared
with the aforementioned alternative methods, this approach can be
practically more advantageous due to its simplicity and tunability.
In its simplest form our modification of the BP iterative procedure
is given by
\begin{eqnarray}
 && \hskip-0.5cm 
\eta_{i\alpha}^{(n+1)}+\frac{1}{\Delta}\sum\limits_{\beta\ni i}
 \eta_{i\beta}^{(n+1)}\label{iter}\\
 &&=h_i+\sum\limits_{\beta\ni i}^{\beta\neq \alpha}
 \tanh^{-1}\left(\prod\limits_{j\in\beta}\tanh\eta_{j\beta}^{(n)}\right)+
 \frac{1}{\Delta}\sum\limits_{\beta\ni i}\eta_{i\beta}^{(n)},
 \nonumber
\end{eqnarray}
where Latin and Greek indexes stand for bits and checks and the
bit/check relations,  e.g. $i\in\alpha$, $\alpha\ni i$  express the
LDPC code considered; $h_i$ is the channel noise-dependent value of
log-likelihoods; and $\eta_{i\alpha}^{(n)}$ is the message
associated at the $n$-th iteration with the edge (of the respective
Tanner graph) connecting $i$-th bit and $\alpha$-th check. $\Delta$
is a tunable parameter. By choosing a sufficiently small $\Delta$
one can guarantee convergence of the iterative procedure to a
minimum of the Bethe free energy.  On the other hand
$\Delta=+\infty$ corresponds exactly to the standard iterative BP.
In the sequel we derive and explain the modified iterative procedure
(\ref{iter}) in detail.

The manuscript is organized as follows. We introduce the Bethe free
energy, the BP equation and the standard iterative BP in Section
\ref{sec:Bethe}. Performance of standard iterative BP, analyzed with a
termination curve, is discussed in Section \ref{sec:Term}. Section
\ref{sec:Relax} describes continuous and sequentially discrete
(iterative) versions of our relaxation method. We discuss performance of
the modified iterative scheme in Section \ref{sec:Perf}, where
Bit-Error-Rate and the termination curve for an LDPC code performed over
Additive-White-Gaussian-Noise (AWGN) channel are discussed for a range
of interesting values of the Signal-to-Noise-Ratios (SNR). We also
discuss here the trade-off between convergence and number of iterations
aiming to find an optimal strategy for selection of the model's
parameters. The last Section \ref{sec:Con} is reserved for conclusions
and discussions.

\section{Bethe Free Energy and Belief Propagation}
\label{sec:Bethe}

Consider a generic factor model \cite{01KFL,01For,04Loe} with a
binary configurational space, $\sigma_i=\pm 1$, $i=1,\cdots,n$,
which is factorized so that the probability $p\{\sigma_i\}$ to find
the system in the state $\{\sigma_i\}$ and the partition function
$Z$ are
\begin{eqnarray}
p\{\sigma_i\}=Z^{-1}\prod\limits_\alpha
f_\alpha(\sigma_\alpha),\quad
Z=\sum\limits_{\{\sigma_i\}}\prod\limits_\alpha
f_\alpha(\sigma_\alpha),
 \label{p1}
\end{eqnarray}
where $\alpha$ labels non-negative and finite factor-functions
$f_\alpha$ with $\alpha=1,\ldots,m$ and $\sigma_\alpha$ represents a
subset of $\sigma_i$ variables. Relations between factor functions
(checks) and elementary discrete variables (bits), expressed as
$i\in\alpha$ and $\alpha\ni i$, can be conveniently represented in
terms of the system-specific factor (Tanner) graph. If $i\in\alpha$
we say that the bit and the check are neighbors. Any spin
(a-posteriori log-likelihood) correlation function can be calculated
using the partition function, $Z$, defined by Eq.~(\ref{p1}).
General expression for the factor functions of an LDPC code is
\begin{eqnarray}
 f_\alpha({\bm \sigma}_\alpha)\equiv\exp
 \left(\sum\limits_{i\in\alpha} h_i\sigma_i/q_i\right)
 \delta\left(\prod\limits_{i\in\alpha}\sigma_i,1\right).
 \label{factor_LDPC}
\end{eqnarray}

Let us now reproduce the derivation of the Belief Propagation equation
based on the Bethe Free energy variational principle, following
closely the description of \cite{05YFW}. (See also the Appendix of
\cite{06CCb}.) In this approach trial probability distributions,
called beliefs, are introduced both for bits and checks $b_i$ and
$b_\alpha$, respectively,  where $i=1,\cdots,N$,
$\alpha=1,\cdots,M$. A belief is defined for given configuration of
the binary variables over the code. Thus, a belief at a bit actually
consists of two probabilities, $b_i(+)$ and $b_i(-)$, and we use a
natural notation $b_i(\sigma_i)$. There are $2^k$ beliefs defined at
a check, $k$ being the number of bits connected to the check, and
we introduce vector notation ${\bm
\sigma}_\alpha=(\sigma_{i_1},\cdots,\sigma_{i_k})$ where
$i_1,\cdots,i_k\in \alpha$ and $\sigma_i=\pm 1$. Beliefs satisfy the
following inequality constraints
\begin{eqnarray}
 0\leq b_i(\sigma_i),b_\alpha({\bm \sigma}_\alpha)\leq 1,
 \label{ineq}
\end{eqnarray}
the normalization constraints
\begin{eqnarray}
 \sum\limits_{\sigma_i}b_i(\sigma_i)= \sum\limits_{{\bm \sigma}_\alpha} b_\alpha({\bm \sigma}_\alpha)=1,
 \label{norm}
\end{eqnarray}
as well as the consistency (between bits and checks) constraints
\begin{eqnarray}
 \sum\limits_{{\bm \sigma}_\alpha\backslash\sigma_i}b_\alpha({\bm
 \sigma}_\alpha)=b_i(\sigma_i),
 \label{cons}
\end{eqnarray}
where ${\bm \sigma}_\alpha\backslash\sigma_i$ stands for the set of
$\sigma_j$ with $j\in \alpha$, $j\neq i$.

The Bethe Free energy is defined as a difference of the Bethe
self-energy and the Bethe entropy,
\begin{eqnarray}
 && \hskip-1.15cm F_{\rm Bethe}=U_{\rm Bethe}-H_{\rm Bethe}, 
\label{Bethe}\\
 && \hskip-1.15cm U_{\rm Bethe}=-\sum\limits_\alpha\sum_{{\bm 
\sigma}_\alpha}b_\alpha({\bm
 \sigma}_\alpha)\ln f_\alpha({\bm \sigma}_\alpha),
 \label{U_Bethe}\\
 && \hskip-1.15cm H_{\rm Bethe}=-\sum\limits_\alpha \sum_{{\bm 
\sigma}_\alpha}b_\alpha({\bm\sigma}_\alpha)
 \ln b_\alpha({\bm\sigma}_\alpha)\nonumber\\ && +\sum\limits_i
 (q_i-1)\sum\limits_{\sigma_i}b_i(\sigma_i)\ln b_i(\sigma_i),
 \label{H_Bethe}
\end{eqnarray}
where ${\bm \sigma}_\alpha=(\sigma_{i_1},\cdots,\sigma_{i_k})$,
$i_1,\cdots,i_k\in \alpha$ and $\sigma_i=\pm 1$. The entropy term
for a bit enters Eq.~(\ref{Bethe}) with the coefficient $1-q_i$ to
account for the right counting of the number of configurations for a
bit: all entries for a bit (e.g. through the check term) should give
$+1$ in total.

Optimal configurations of beliefs are the ones that minimize the
Bethe Free energy (\ref{Bethe}) subject to the constraints
(\ref{ineq},\ref{norm},\ref{cons}). Introducing these constraints
into the effective Lagrangian through Lagrange multiplier terms
\begin{eqnarray}
 && \hskip-0.35cm L=F_{\rm Bethe}+
 \sum\limits_\alpha
 \gamma_\alpha \left(\sum\limits_{{\bm \sigma}_\alpha}
 b_\alpha({\bm \sigma}_\alpha)-1\right)\label{Lagr}\\ && +
 \sum\limits_{i}\gamma_i
 \left(\sum\limits_{\sigma_i}b_i(\sigma_i)-1\right)\nonumber\\ && +
 \sum\limits_i\sum\limits_{\alpha\ni i}\sum\limits_{\sigma_i}
 \lambda_{i\alpha}(\sigma_i)\left(b_i(\sigma_i)-
 \sum\limits_{{\bm \sigma}_\alpha\backslash\sigma_i}b_\alpha({\bm
 \sigma}_\alpha)\right),\nonumber
\end{eqnarray}
and looking for the extremum with respect to all possible beliefs
leads to
\begin{eqnarray}
 && \!\!\!\!\!\! \frac{\delta L}{\delta b_a({\bm \sigma}_a)} 
  = 0 \label{Lba} \\
  && \!\!\Rightarrow\quad
 b_\alpha({\bm \sigma}_\alpha)=f_\alpha({\bm\sigma}_\alpha)
 \exp\left[-\gamma_\alpha-1+\sum\limits_{i\in\alpha}\lambda_{i\alpha}(\sigma_i)\right],
 \nonumber\\
 && \!\!\!\!\!\! \frac{\delta L}{\delta b_i(\sigma_i)} = 0 \label{Lbi} 
   \\
 && \!\!\Rightarrow\quad
 b_i(\sigma_i)=\exp\left[\frac{1}{q_i-1}\left(\gamma_i+
 \sum\limits_{\alpha\ni i}\lambda_{i\alpha}(\sigma_i)\right)-1\right].
 \nonumber
\end{eqnarray}
Substituting $\lambda_{i\alpha}(\sigma_i)\equiv\ln\prod_{\beta\ni i;
\beta\neq\alpha}\mu_{i\beta}(\sigma_i)$ into
Eq.(\ref{Lba},\ref{Lbi}) we arrive at
\begin{eqnarray}
 && \hskip-1.1cm b_\alpha({\bm\sigma}_\alpha)\propto f_\alpha({\bm 
\sigma}_\alpha)\prod\limits_{i\in\alpha}
 \prod\limits_{\beta\ni i}^{\beta\neq\alpha}
 \mu_{i\beta}(\sigma_i)\nonumber\\ && =
 f_\alpha({\bm \sigma}_\alpha)\prod\limits_{i\in\alpha}
 \exp\left(\lambda_{i\alpha}(\sigma_i)\right)
 ,\label{ba}\\
&& \hskip-1.1cm b_i(\sigma_i)\propto \prod\limits_{\alpha\ni i}
 \mu_{i\alpha}(\sigma_i)=
 \exp\left(\frac{\sum_{\alpha\ni i} \lambda_{i\alpha}(\sigma_i)}{q_i-1}\right),
\label{bi}
\end{eqnarray}
where $\propto$ is used to indicate that we should use the
normalization conditions (\ref{norm}) to guarantee that the beliefs
sum up to one. Applying the consistency constraint (\ref{cons}) to
Eqs.~(\ref{ba}), making summation over all spins but the given
$\sigma_i$, and also using Eq.~(\ref{bi}) we derive the following BP
equations
\begin{eqnarray}
 && \!\!\!\!\!\!\!\! \prod\limits_{\alpha\ni 
i}\mu_{i\alpha}(\sigma_i)\propto
 b_i(\sigma_i)\label{ba0}\\ && \propto \left(\prod\limits_{\beta\ni
 i}^{\beta\neq\alpha} \mu_{i\beta}(\sigma_i)\right)
 \sum\limits_{\sigma_\alpha\setminus\sigma_i}f_\alpha(\sigma_\alpha)
 \prod\limits_{j\in\alpha}^{j\neq
 i}\prod\limits_{\beta\ni j}^{\beta\neq
 \alpha}\mu_{j\beta}(\sigma_j).
 \nonumber
\end{eqnarray}
The right hand side of Eq.~(\ref{ba0}) rewritten for the LDPC case (\ref{factor_LDPC}) becomes
\begin{eqnarray}
 && \!\!\!\!\!\!\!\! b_i(\sigma_i)\propto 
\exp(h_i\sigma_i)\left(\prod\limits_{\beta\ni
 i}^{\beta\neq\alpha} \mu_{i\beta}(\sigma_i)\right)\label{ba1}\\ &&
 \!\!\!\!\!\!\!\!\times\left(\prod\limits_{j\in\alpha}^{j\neq
 i}\!\left(\mu_{j\alpha}(+)\!+\!\mu_{j\alpha}(-)\right)\!+\!\sigma_i
 \prod\limits_{j\in\alpha}^{j\neq
 i}\left(\mu_{j\alpha}(+)\!-\!\mu_{j\alpha}(-)\right)\right).
 \nonumber
\end{eqnarray}
Thus constructing $b_i(+)/b_i(-)$ for the LDPC case in two different
ways (correspondent to left and right relations in Eq.~(\ref{ba0})),
equating the results and introducing the $\eta_{i\alpha}$ field
\begin{eqnarray}
\exp(2\eta_{i\alpha})=\frac{\mu_{i\alpha}(+)}{\mu_{i\alpha}(-)},
\label{eta_b}
\end{eqnarray}
one arrives at the following BP equations for the $\eta_{i\alpha}$
fields:
\begin{eqnarray}
 \eta_{i\alpha}=h_i+\sum\limits_{\beta\ni i}^{\beta\neq \alpha}
 \tanh^{-1}\left(\prod\limits_{j\in\beta}^{j\neq
 i}\tanh\eta_{j\beta}\right).
 \label{bp1}
\end{eqnarray}
Iterative solution of this equation corresponding to
Eq.~(\ref{iter}) with $\Delta=+\infty$ is just a standard iterative
BP (which can also be called sum-product) used for the decoding of an
LDPC code.

A simplified min-sum version of Eq.~(\ref{iter}) is
\begin{eqnarray}
 && \!\!\!\!\!\!\!\! 
\eta_{i\alpha}^{(n+1)}+\frac{1}{\Delta}\sum\limits_{\beta\ni i}
 \eta_{i\beta}^{(n+1)}\label{min-sum}\\
 &&=h_i+
\sum_{\beta\neq\alpha}^{\beta\ni i}
  \prod_{j\neq i}^{j\in\beta} \mbox{sign}
    \big[ \eta^{(n)}_{j\beta} \big] \min_{j\neq i}^{j\in\beta} \big|
    \eta^{(n)}_{j\beta} \big|+
 \frac{1}{\Delta}\sum\limits_{\beta\ni i}\eta_{i\beta}^{(n)},
 \nonumber
\end{eqnarray}

\section{Termination curve for standard iterative BP}
\label{sec:Term}

To illustrate the standard BP iterative decoding, given by
Eqs.~(\ref{iter},\ref{min-sum}) with $\Delta = +\infty$, we consider the
example of the $[155,64,20]$ code of Tanner \cite{01TSF} performing over
AWGN channel channel characterized by the transition probability for a
bit, $p(x|\sigma) = \exp(-s^2 (x - \sigma)^2/2)/\sqrt{2\pi/s^2}$, where
$\sigma$ and $x$ are the input and output values at a bit and $s^2$ is
the SNR. Launching a fixed codeword into the channel, emulating the
channel by means of a standard Monte-Carlo simulations and then decoding
the channel output constitutes our ``experimental" procedure.

\begin{figure}[ht]
  \centerline{\includegraphics[width=3in]{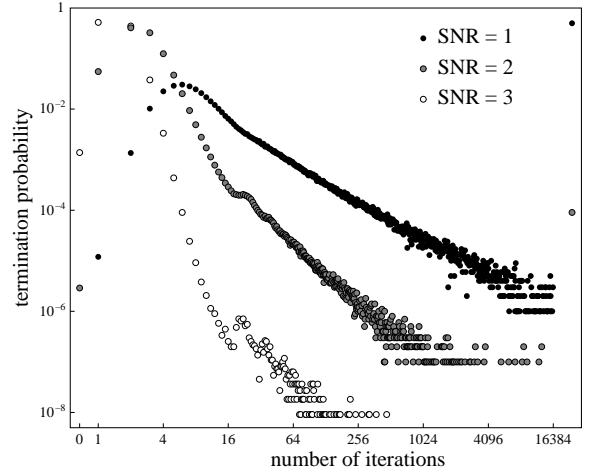}}
  \caption{The termination probability curve for $\mbox{SNR} = 1, 2,
  3$. Notice that the probability of termination (successful decoding)
  without any iterations is always finite.
  Few points on the right part of the plot correspond to the
  case when the decoding was not terminated even at
  the maximum number of iterations, $16384$ (decoding fails to converge to a codeword).
  \label{TC123}}
\end{figure}

We analyze the probability distribution function of the iteration number
$n_{\rm it}$ at which the decoding terminates. The termination
probability curve for the standard min-sum, described by
Eq.~(\ref{min-sum}) with $\Delta=+\infty$, is shown in Fig.~\ref{TC123}
for $\mbox{SNR} = 1, 2, 3$.

The result of decoding is also verified at each iteration step for
compliance with a codeword: iteration is terminated if a codeword is
recovered. This termination strategy can still give an error, although
the probability to confuse actual and a distant codewords is much less
than the probability not to recover a codeword for many iterations. If
one neglects the very low probability of the codewords' confusion, then
the probability of still having a failure after $n_{\rm it}$ iterations
is equal to the integral/sum over the termination curve from $n_{\rm
it}$ and up. Note also that the probability that even infinite number of
iterations will not result in a codeword can actually be finite.

Discussing Fig.~\ref{TC123} one observes two distinct features of the
termination probability curve. First, in all cases the curve reaches its
maximum at some relatively small number of iterations. Second, each
curve crosses over to an algebraic-like decay which gets steeper with
the SNR increase.

The emergence of an algebraically extended tail (that is a tail which
does not decay fast) is not encouraging, as it suggests that increasing
the number of iteration will not bring much of an improvement in the
iterative procedure. It also motivates us to look for possibilities of
accelerating convergence of the BP algorithm to a minimum of the Bethe
free energy.

Note also the wiggling of termination curves for ${\rm SNR} = 2, 3$ near
the crossover point (see Fig.~\ref{TC123}). It is possibly related to
the cycling of the BP dynamics (and thus the inability of BP to
converge).

\section{Relaxation to a minimum of the Bethe free energy}
\label{sec:Relax}

The idea is to introduce relaxational dynamics (damping) in an
auxiliary time, $t$, thus enforcing convergence to a minimum of the
Bethe Free energy. One chooses $b_i(\sigma_i)$ as the main
variational field and considers relaxing  variational equations
Eqs.~(\ref{Lbi}) according to
\begin{eqnarray}
\partial_t b_i(\sigma_i)=-\frac{1}{\tau_i}\frac{\delta L}{\delta
b_i(\sigma_i)}, \label{relax_bi}
\end{eqnarray}
while keeping the set of remaining variational equations
Eqs.~(\ref{norm},\ref{Lba},\ref{cons}) intact. Here positive
parameters $\tau_i$ have the physical meaning of
correlation/relaxation times. Performing calculations, that are
completely equivalent to the ones described in Section
\ref{sec:Bethe}, we arrive at the following modified BP equations
\begin{eqnarray}
 && \!\!\!\!\!\!\!\!\!\!\!\!\eta_{i\alpha}\!+\!(q_i\!-\!1)Q_i\!
 =\!h_i\!+\!\sum\limits_{\beta\ni i}^{\beta\neq \alpha}
 \tanh^{-1}\left(\prod\limits_{j\in\beta}^{j\neq
 i}\tanh\eta_{j\beta}\right),
 \label{bp2}\\
 && Q_i=\tau_i\partial_t \tanh\left(\sum\limits_{\alpha\ni
 i}\eta_{i\alpha}+(q_i-1)Q_i\right).\label{Qeq}
\end{eqnarray}
We are interested to approach (find) a solution of the original BP
Eq.~(\ref{bp1}). One assumes $Q_i\ll\sum_{\alpha\ni
 i}\eta_{i\alpha}$,  thus ignoring the second term under $\tanh$ in
 Eq.~(\ref{Qeq}). The resulting continuous equation is
\begin{eqnarray}
 && \hskip-1cm 
\eta_{i\alpha}+\frac{(q_i-1)\tau_i}{\cosh^2(\sum_{\alpha\ni 
i}\eta_{i\alpha})}
 \partial_t\left(\sum\limits_{\alpha\ni i}\eta_{i\alpha}\right)\nonumber\\
 &&=h_i+
 \sum\limits_{\beta\ni i}^{\beta\neq \alpha}
 \tanh^{-1}\left(\prod\limits_{j\in\beta}^{j\neq
 i}\tanh\eta_{j\beta}\right).
 \label{bp3}
\end{eqnarray}
Eq.~(\ref{iter}) represents a simple discretized version of the
Eq.~(\ref{bp3}) where the correlation coefficients $\tau_i$ are
chosen to make the coefficient in front of the second term on the
left hand side of Eq.~(\ref{bp3}) independent of the bit index, $i$.
Then the resulting time dependent coefficient can be rescaled to one
by an appropriate choice of the temporal unit; $t_n$ is the uniform
discrete time, $n$ is positive integer, $t_{n+1}-t_n=\Delta>0$; the
left hand side (right hand side) of Eq.(\ref{bp3}) is taken at
$t_{n+1}$ ($t_n$) and the temporal derivative is discretized in a
standard retarded way, $\partial_t\eta_{i\alpha}\to
(\eta_{i\alpha}^{(n+1)}-\eta_{i\alpha}^{(n)})/\Delta$. This choice
of relaxation coefficients and discretization, resulted in
Eq.~(\ref{iter}), was taken out of consideration in the final
formula for simplicity, realizability at all positive $\Delta$ and also
its equivalence to the standard iterative BP at $\Delta\to+\infty$.




\section{Modified iterative BP: test of performance}
\label{sec:Perf}

\begin{figure}[ht]
  \centerline{\includegraphics[width=3in]{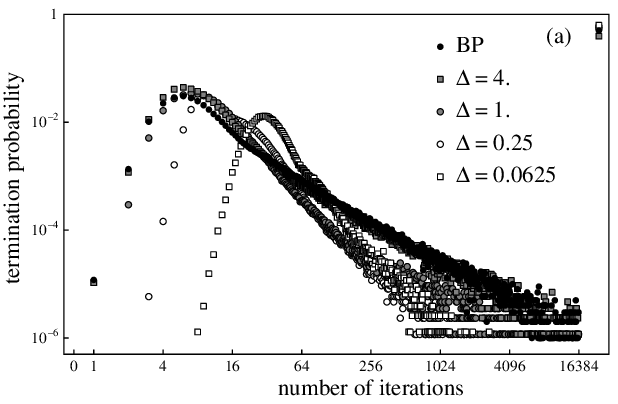}}
  \smallskip
  \centerline{\includegraphics[width=3in]{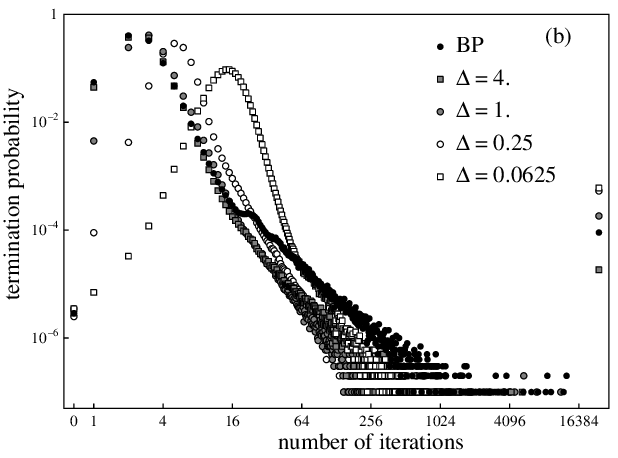}}
  \smallskip
  \centerline{\includegraphics[width=3in]{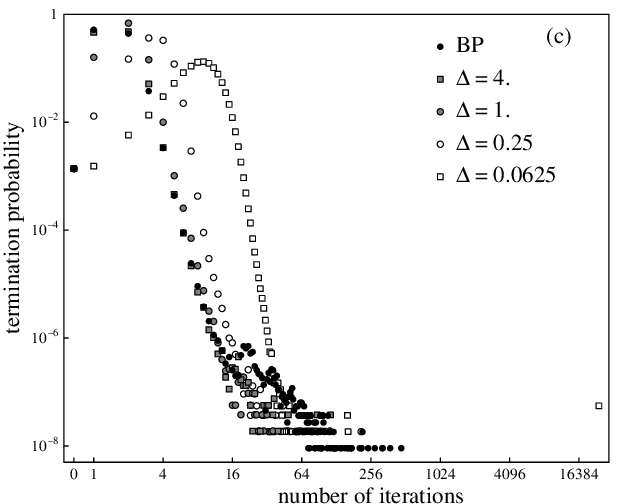}}
  \caption{The termination probability curves for $\mbox{SNR} = 1$ (a), 
    $\mbox{SNR} = 2$ (b), and $\mbox{SNR} = 3$ (c).
    \label{TC}}
\end{figure}

We test the min-sum version (\ref{min-sum}) of the modified iterative BP
with the Monte Carlo simulations of the $[155,64,20]$ code at few values
of SNRs.  The resulting termination probability curves are shown in
Fig.~\ref{TC} for ${\rm SNR} = 1, 2, 3$.

The simulations show a shift of the probability curve maximum to the
right (towards larger number of iterations) with the damping parameter
decrease however once the maximum is achieved, the decay of the curve at
a finite $\Delta$ is faster with the number of iterations than in the
standard BP case. The decay rate actually increases as $\Delta$
decreases.

We conclude that at the largest $n_{\rm it}$ the performance of a
modified iterative BP is strictly better. However to optimize the
modified iterative BP, thus aiming at better performance than given by
the standard iterative BP, one needs to account for the trade-off
between decreasing $\Delta$ leading to a faster decay of the termination
probability curve at the largest $n_{\rm it}$, but on the other side it
comes with the price in the actual number of iteration necessary to
achieve the asymptotic decay regime.

\begin{figure}[ht]
  \centerline{\includegraphics[width=3in]{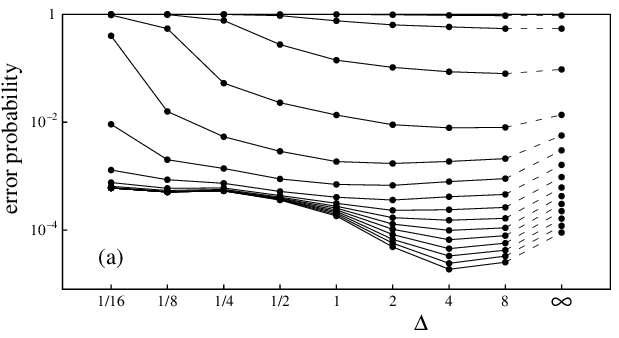}}
  \smallskip
  \centerline{\includegraphics[width=3in]{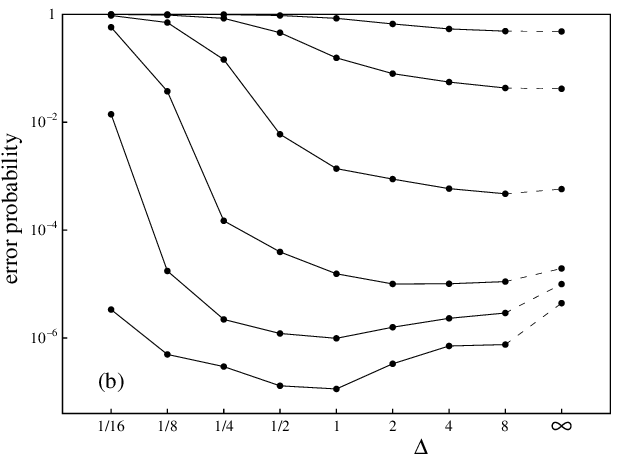}}
  \caption{Decoding error probability as a function of $\Delta$ for
    $\mbox{SNR} = 2$ (a) and $\mbox{SNR} = 3$ (b). Different curves 
    correspond to different maximally allowed $n_{\rm it}$: starting 
    from $n_{\rm it} = 1$ (top curve) and increasing $n_{\rm it}$ by 
    factor of $2$ with each next lower curve. The points on the right 
    correspond to the standard BP ($\Delta = \infty$).
  \label{ER}}
\end{figure}

The last point is illustrated by Fig.~\ref{ER}, where the decoding error
probability depends non-monotonically on $\Delta$. One can also see that
the modification of BP could improve the decoding performance; e.g., at
${\rm SNR} = 3$ and maximally allowed $n_{\rm it} = 32$ (after which the
decoding unconditionally stops) the decoding error probability is
reduced by factor of about 40 by choosing $\Delta = 1$ (see the bottom
curve at Fig.~\ref{ER}(b)).

\section{Conclusions and Discussions}
\label{sec:Con}

We presented a simple extension of the iterative BP which allows (with
proper optimization in the $\Delta-n_{\rm it}$ plane) to guarantee not
only an asymptotic convergence of BP to a local minimum of the Bethe
free energy but also a serious gain in decoding performance at finite
$n_{\rm it}$.

In addition to their own utility, these results should also be useful
for systematic improvement of the BP approximation. Indeed, as it was
recently shown in \cite{06CCa,06CCb} solution of the BP equation can be
used to express the full partition function (or a-posteriori
log-likehoods calculated within MAP) in terms of the so-called loop
series, where each term is associated with a generalized loop on the
factor graph. This loop calculus/series offers a remarkable opportunity
for constructing a sequence of efficient approximate and systematically
improvable algorithms. Thus we anticipate that the improved iterative BP
discussed in the present manuscript will become an important building
block in this future approximate algorithm construction.

We already mentioned in the introduction that our algorithm can be
advantageous over other BP-based algorithms converging to a minimum of
the Bethe free energy mainly due to its simplicity and tunability. In
particular, the concave-convex algorithms of \cite{02Yui,03HAK}, as well
as related linear programming decoding algorithms \cite{03FWK}, are
formulated in terms of beliefs. On the contrary our modification of the
iterative BP can be extensively simplified and stated in terms of the
fewer number of $\eta$ fields each associated with an edge of the factor
graph rather than with much bigger family of local code-words. Thus in
the case of a regular LDPC code with $M$ checks of the connectivity
degree $k$ one finds that the number of variables taken at each step of
the iterative procedure is $k*M$ and $2^{k-1}*M$ in our iterative scheme
and in the concave-convex scheme respectively. Having a tunable
correlation parameter $\tau$ in the problem is also advantageous as it
allows generalizations (e.g. by turning to a individual bit dependent
relaxation rate). This flexibility is particularly desirable in the
degenerate case with multiple minima of the Bethe free energy, as it
allows a painless implementation of annealing as well as other more
sophisticated relaxation techniques speeding up and/or improving
convergence.


\begin{thebibliography}{99}

\bibitem{63Gal} R.G. Gallager, {\it Low density parity check
codes} (MIT PressCambridhe, MA, 1963).

\bibitem{99Mac} D.J.C.~MacKay, {\it Good error-correcting codes based on
very sparse matrices}, IEEE Trans. Inf. Theory~{\bf 45} (2) 399-431
(1999).

\bibitem{03RU} T.~Richardson, R.~Urbanke,
{\it The renaissance of Gallager's low-density parity-check codes},
{IEEE Communications Magazine}~{\bf 41}, 126--131 (2003).


\bibitem{05YFW} J.S. Yedidia, W.T. Freeman, Y. Weiss,
{\it Constructing Free Energy Approximations and Generalized Belief
Propagation Algorithms},
IEEE IT{\bf 51}, 2282 (2005).

\bibitem{88Pea} J. Pearl, {\it Probabilistic reasoning in intelligent
systems: network of plausible inference} (Kaufmann, San Francisco,
1988).

\bibitem{02TJ} S.~Tatikonda, M.I.~Jordan, {\it Uncertainty in
Artificial Intelligence: Proceedings of the 18th
conference(UAI-2002)} (San Francisco: Morgan Kaufmann Publishing) p.
493.

\bibitem{04Hes} T. Heskes, {\it On the Uniqness of Loopy Belief
Propagation Fixed Points}, Neural Computation {\bf 16}, 2379-2413
(2004).

\bibitem{03WJW} M.J.~Wainwright, T.S. Jaakola, A.S.~Willsky, {\it
Tree-Based Reparamterizations Framework for Analysis of Sum-Product
and Related Algorithms}, IEEE IT {\bf 49}, 1120-1146 (2003).

\bibitem{02Yui} A.L.~Yuille, {\it CCCP algorithms to minimize the
Bethe and Kikuchi free energies: convergent alternatives to belief
propagation}, Neural Computation {\bf 14}, 1691-1722 (2002).

\bibitem{03HAK} T.~Heskes, K.~Albers, B.~Kappen, {\it Uncertainty in
artificial intelligence: Proceedings of the 19th
conference(UAI-2003)} (San Francisco: Morgan Kaufmann Publishing) p.
313.

\bibitem{01KFL} F.~R.~Kschischang, B.~J.~Frey, and H.-A.~Loeliger,
{\it Factor gaphs and the sum-product algorithm}, IEEE IT {\bf 47},
498-519 (2001).

\bibitem{01For} G.~D.~Forney, {\it Codes on graphs: normal 
realizations},
IEEE IT {\bf 47}, 520-548 (2001).

\bibitem{04Loe} H.-A.~Loeliger, {\it An Introduction to factor graphs},
IEEE Signal Processing Magazine, Jan 2001, p.~28-41.

\bibitem{01TSF} R.M.~Tanner, D.~Srkdhara, T.~Fuja,
{\it A class of group-structured LDPC codes}, Proc. of ICSTA 2001,
Ambleside, England.

\bibitem{06CCa} M. Chertkov, V. Chernyak, {\it Loop calculus in 
statistical physics and information science},  Phys. Rev. E {\bf 73}, 
065102(R) (2006); cond-mat/0601487.

\bibitem{06CCb} M. Chertkov, V. Chernyak,
{\it Loop series for discrete statistical models on graphs}, J.
Stat. Mech. (2006) P06009,
cond-mat/0603189. 

\bibitem{03FWK} J.~ Feldman, M.~ Wainwright,  D.R.~Karger, {\it
Using linear programming to decode linear codes}, 2003 Conference on
Information Sciences and Systems,  The John Hopkins University,
March 12-14, 2003.

\end{thebibliography}
\end{document}